\documentclass[superscriptaddress,amsmath,amssymb,aps,prl,longbibliography,twocolumn]{revtex4-2}

\usepackage{mathrsfs}
\usepackage{amssymb}
\usepackage{amsmath}
\usepackage{graphicx}
\usepackage{subfigure}
\usepackage{dcolumn}
\usepackage{url}
\usepackage{color}
\usepackage[colorlinks=true,
            linkcolor=red,
            citecolor=blue,
            urlcolor=blue]{hyperref}
\usepackage{paralist}
\usepackage{ulem}
\usepackage{soul}

\newcommand{\sgn}{\mathrm{sgn}}

\begin{document}

\title{Quantum droplets in one-dimensional mixtures of quasi Bose-Einstein condensates and Tonks-Girardeau gases}  

\author{Wen-Bin He}
\email{wenbin.he@oist.jp}
\affiliation{Quantum Systems Unit \& Okinawa Center for Quantum Technologies, Okinawa Institute of Science and Technology Graduate University, 904-0495 Okinawa, Japan}%
\affiliation{School of Mathematics and Physics, Jinggangshan University, Ji’an 343009, China }

\author{Su Yi}
\email{yisu@nbu.edu.cn}  
\affiliation{Institute of Fundamental Physics and Quantum Technology \& School of
Physical Science and Technology, Ningbo University, Ningbo, 315211, China}

\author{Thomas Busch}%
\email{thomas.busch@oist.jp}
\affiliation{Quantum Systems Unit, Okinawa Institute of Science and Technology Graduate University, 904-0495 Okinawa, Japan}%

\date{\today}

\begin{abstract}
    While binary atomic Bose-Einstein condensates are typically prone to collapse under strong interspecies attraction, it has been shown that higher-order fluctuation corrections, known as Lee-Huang-Yang corrections, can stabilize the mixture. In this work, we demonstrate an alternative stabilization mechanism based on kinetic energy. Specifically, we consider a one-dimensional mixture of a quasi-BEC and a Tonks–Girardeau gas, and show that the kinetic energy of the TG component can counteract the interspecies attraction, thereby preventing collapse. This balance leads to the formation of a self-bound quantum droplet, which exhibits two distinct regimes: a low-density and a high-density droplet. We argue that these regimes are smoothly connected by a crossover. Furthermore, an analysis of the derivatives of the ground state energy and bulk modulus reveals that the transition from a miscible mixture to the droplet phase is of third order. Our findings extend the theoretical understanding of quantum droplets in low-dimensional quantum gases, and the proposed system may be experimentally accessible within current ultracold atom platforms.
\end{abstract}

\maketitle

\section{ Introduction}

Quantum droplets (QDs) in ultracold atomic gases represent an unusual state of matter characterized by their self-bound nature that ensures stability even in the absence of external confinement. In binary Bose-Einstein condensates (BECs), D.S.~Petrov \cite{Petrov} demonstrated that attractive mean-field interactions can be stabilized against collapse by quantum fluctuations, specifically through higher-order Lee-Huang-Yang (LHY) corrections \cite{LHY}. This necessitates a beyond-mean-field theoretical treatment of the system, and as such, the study of quantum droplets has attracted significant theoretical attention in recent years. Various theoretical approaches have been employed to explore the properties of LHY-stabilized quantum droplets, including bosonic pairing models for self-bound states \cite{huliu}, investigations into their collective excitations \cite{Tylutki,huliu_pra}, and studies of quantum criticality in these systems \cite{Spada,zqyu23}, quantum droplet in Bose-Fermi mixture \cite{rakshit2019} and Bose-Bose mixture. Furthermore, long-range interactions have been shown to give rise to a diverse array of quantum droplet phenomena in dipolar quantum gases \cite{Igor,Natale,Bisset,chomaz2022dipolar,yongyao,yongyao_review}. The existence of quantum droplets has also been confirmed experimentally by several groups \cite{Cabrera,exp_Semeghini,exp_Errico}.

The original theoretical framework for quantum droplets showed that a negative mean-field interaction energy term, scaling as $n^2$ (where $n$ is the particle density), can be counterbalanced by a positive LHY correction term scaling as $n^{5/2}$. This faster growth of the LHY term with density enables the formation of a self-bound state as the system size increases \cite{Petrov,Ota_sci,Skov_lhy,boudjemaa}. For convenience, we refer to such droplets as LHY quantum droplets. The transition from a miscible mixture to a quantum droplet can be estimated from a mean-field stability criterion, $|g_{AB}| > \sqrt{g_{A} g_{B}}$ \cite{Petrov}, where $g_A$ and $g_B$ denote the intra-species interaction strengths, and $g_{AB}$ represents the inter-species attraction. However, this mechanism requires precise control of higher-order LHY excitations, which imposes considerable experimental challenges in realizing and probing these droplets.

In this work, we propose an alternative mechanism for stabilizing quantum droplets in bosonic mixtures, based on kinetic energy contributions rather than fluctuation-induced corrections. Specifically, we consider a one-dimensional mixture composed of a weakly interacting quasi-Bose-Einstein condensate (BEC) and a strongly interacting Tonks-Girardeau (TG) gas. In such systems, the kinetic energy of the TG component exhibits a cubic scaling with density, $\langle \hat{T}_{TG} \rangle \propto n^{3}$, which is significantly steeper than the quadratic scaling characteristic of mean-field interactions. This unique feature of TG gases, which has been experimentally observed in previous studies \cite{kinoshita2004,paredes2004tonks}, enables a natural stabilization of the mixture against collapse induced by attractive inter-species interactions. Beyond mere stabilization, the interplay between the BEC and TG components gives rise to quantum droplets with intriguing properties that distinguish them from previously studied LHY-stabilized droplets.  The fact that the stabilization mechanism relies on the cubic kinetic energy balancing the quadratic mean-field interaction between the BEC and TG components leads to a quantum mixture which displays a crossover from low-density to high-density droplets, and the appearance of a higher-order phase transition.

In the following, we present a detailed analysis of the formation and  properties of quantum droplets in a one-dimensional mixture of a weakly interacting Bose-Einstein condensate and a strongly interacting Tonks-Girardeau gas. We begin by introducing the coupled equations of motion (EOMs) that describe the properties of the BEC-TG mixture, followed by a derivation of the corresponding energy functional and its associated critical point for droplet formation. Through exact numerical solutions of the EOMs, we obtain the ground-state density distributions of the BEC and TG components as a function of the inter-species attraction strength,  $g_{\mathrm{int}}$.

Our results reveal the emergence of a self-bound quantum droplet, characterized by a localized BEC density profile with a flat-top structure, and a TG gas density that exhibits both a central hump and a finite delocalized background. By quantifying the density overlap length and the central BEC density, we distinguish between two regimes: a low-density and a high-density quantum droplet, which are connected by a crossover. We further show that the transition from a miscible mixture to the quantum droplet state is a third-order phase transition, which is markedly different from the first-order transition associated with LHY-stabilized droplets in Bose-Bose mixtures. The nearly convex behavior of the ground-state energy as a function of total density in the balanced case supports the energetic stability of the droplet. This different paradigm for droplet formation in low-dimensional quantum fluids open the door for more complex theoretical and experimental investigations into strongly correlated quantum mixtures.

\begin{figure*}[tb]
    \includegraphics[scale=0.32]  {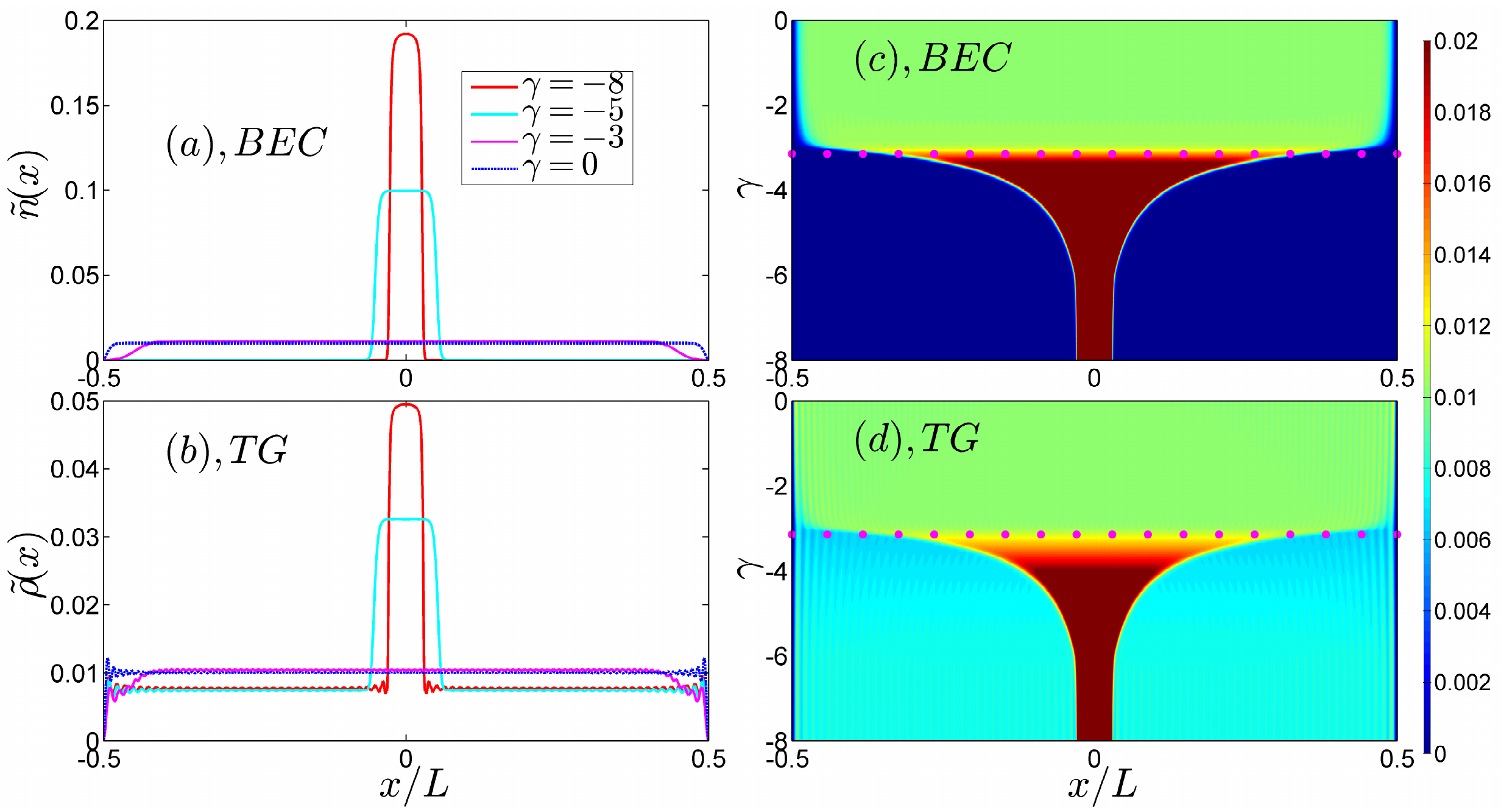}
    \caption{Ground state  rescaled density distribution of (a) the BEC  $\tilde{n}(x)=n(x)/N_{BEC}/n_{F}$ and (b) the TG gas $\tilde{\rho}(x)=\rho(x)/N_{TG}/n_{F}$ v.s $\tilde{x}=x/L$ for different values the  rescaled attractive interaction strength $\gamma$. The systems size $N=100$ and $\tilde{g}_\text{BEC}=1$. Panels (c) and (d) show  contour plots of the density distribution of the BEC and the TG  gas, respectively, as a function of the attractive inter-component interaction strength. For subplots(c) and (d), in order to show the quantum droplet clearly, we set colorbar to be the same for the BEC and the TG gas with the maximum about twice the value of the averaged normalized density $\tilde{n}\sim 0.01$.  The transition to the droplet state can be seen and the critical point obtained with the MF method is indicated by the horizontal magenta dots.
     }
    \label{tgbec_att_nbec100ntg100L100_gbec1}
\end{figure*}

\section{Model } 
We model the one-dimensional mixture of a weakly interacting Bose-Einstein condensate (BEC) and a strongly correlated Tonks–Girardeau (TG) gas by coupling the mean-field Gross–Pitaevskii equation (GPE) for the BEC and the Schrödinger equation for the TG gas via  density–density interaction terms \cite{Keller:20}. While solving many-body Schrödinger equations is generally challenging, the TG limit of infinitely strong intra-component repulsion allows one to employ the Bose–Fermi mapping theorem \cite{Girardeau}, which maps the bosonic system onto a system of non-interacting fermions. For a simpler discussion of our results, we assume in the following equal masses for the BEC and TG components, $m_\text{BEC} = m_\text{TG} = m$. The energy functional of the system,  $E(\psi(x), \phi_n(x))$, is then given by
\begin{align}
    E=&\int dx\; \left(  \psi^*(x) \left[-\frac{\hbar^2}{2m}\frac{\partial^2}{\partial x^2}+\frac{1}{2}g_\text{BEC}|\psi|^2\right]\psi(x) \right. \nonumber \\
    & \left. +\sum_{n=1}^{N_{TG}}\left[- \frac{\hbar^2}{2m}\phi_n^{*}(x)\frac{\partial^2}{\partial x^2}\phi_n(x)\right]+  g_\text{int}|\psi(x)|^2 \rho(x) \right) , 
    \label{eq_E}
\end{align}
where  $\psi(x)$  is the mean-field wavefunction of the condensate and  the $\phi_n(x)$  are the single-particle states required to calculate the TG gas wavefunction $\Phi(x_1,\cdots,x_{N_{TG}})=\prod_{i<j}\sgn(x_i-x_j) \det_{m,n}[\phi_{m}(x_{n})]/\sqrt{N_{TG}!}$. 
The first two terms represent the standard kinetic and nonlinear interaction energies of the BEC and the kinetic energy of an $N_{TG}$ particle TG gas, respectively. The third term accounts for the interspecies interaction between the two components, modeled as a density–density coupling, where the TG gas density is given by  $\rho(x) = \sum_{n=1}^{N_{TG}} |\phi_n(x)|^2$. The interspecies interaction is given by $g_\text{int}<0$ and the interaction within the BEC by $g_\text{BEC}>0$. This $s$-wave approximation describes scattering in ultracold atomic experiments well \cite{bloch}, however the inclusion of higher-order partial-wave interactions \cite{feshbach} in one dimension can introduce small corrections, which would lead to slight shifts in the transition point of the quantum mixture. In writing this functional we have also assumed that both gases are confined in a box potential of size $L$, so that the density distribution of the BEC, $n(x)=|\psi(x)|^2$, and the TG gas satisfy the normalization condition
\begin{equation}
 \int_L dx\; n(x)=N_\text{BEC},\qquad  \int_L dx \rho(x) =N_{\text{TG}}.
\end{equation}
Using the Lagrange multiplier method we get the action functional
\begin{equation}
    S=\int \mathcal{L}(x) dx=E-\mu \int dx |\psi(x)|^2- \sum_{n}\epsilon_{n} \int dx |\phi_n(x)|^2,
\end{equation}
which should be extremised by the wave functions $\psi(x)$ and $\{ \phi_n(x) \}$ as
\begin{equation}
    \frac{\delta \mathcal{L}(x)}{\delta \psi^{*}(x)}=0\quad\text{and}\quad \frac{\delta \mathcal{L}(x)}{\delta \phi_n^{*}(x)}=0.
\end{equation}
By minimizing above functional, one can derive the coupled equations of motion for the two components of the mixture as \cite{Karpiuk}
\begin{align} 
    \left(-\frac{\hbar^2}{2m}\frac{\partial^2}{\partial x^2}+ g_\text{BEC}|\psi|^2 +g_\text{int}\rho(x) \right)\psi(x)&=\mu\psi(x), \label{eq_bec} \\
    \left(-\frac{\hbar^2}{2m}\frac{\partial^2}{\partial {x^2}}+g_\text{int} |\psi(x)|^2\right)\phi_{n}(x)&=\epsilon_{n} \phi_{n}(x).
 \label{eq_tg}
\end{align}

In the following we will solve the coupled equations of motion for the ground state of the BEC-TG mixture, Eqs.~\eqref{eq_bec} and \eqref{eq_tg}, using imaginary time evolution and iteration until convergence. While this model is known to predict strong localisation in highly population imbalanced systems, recent studies have shown that this is blocked by the presence of correlations between the two components, which are neglected in the mean-field approach \cite{Filatrella,GomezLozada:25,Breu:25}. In this work we therefore always consider the case where the two particle numbers are of the same order $N_{BEC}=N_{TG}=N$ and unity density $n=N/L=1$. To help grasp the physical implications, we scale all quantities in the figures in terms of the Fermi energy $E_{F}=\frac{\hbar^2 k_{F}^2}{2m}$ and  momentum $k_{F}=\pi n_{F}$,  and the density $n_{F}=\frac{N}{L}$ of the TG gas. With this, the system parameters in the energy functional are given by
\begin{align*}
     &g_{BEC}/n_{F} \rightarrow  \tilde{g}_{BEC}, \\
     &g_{int}/n_{F} \rightarrow  \gamma, \\
     & E/N/E_{F} \rightarrow \tilde{\epsilon}, \\
      & x/L \rightarrow \tilde{x}  \nonumber
\end{align*}
where the average energy of a single particle is given by $\epsilon=E/N$.

\begin{figure}[tb]
    \subfigure{ \includegraphics[scale=0.375]{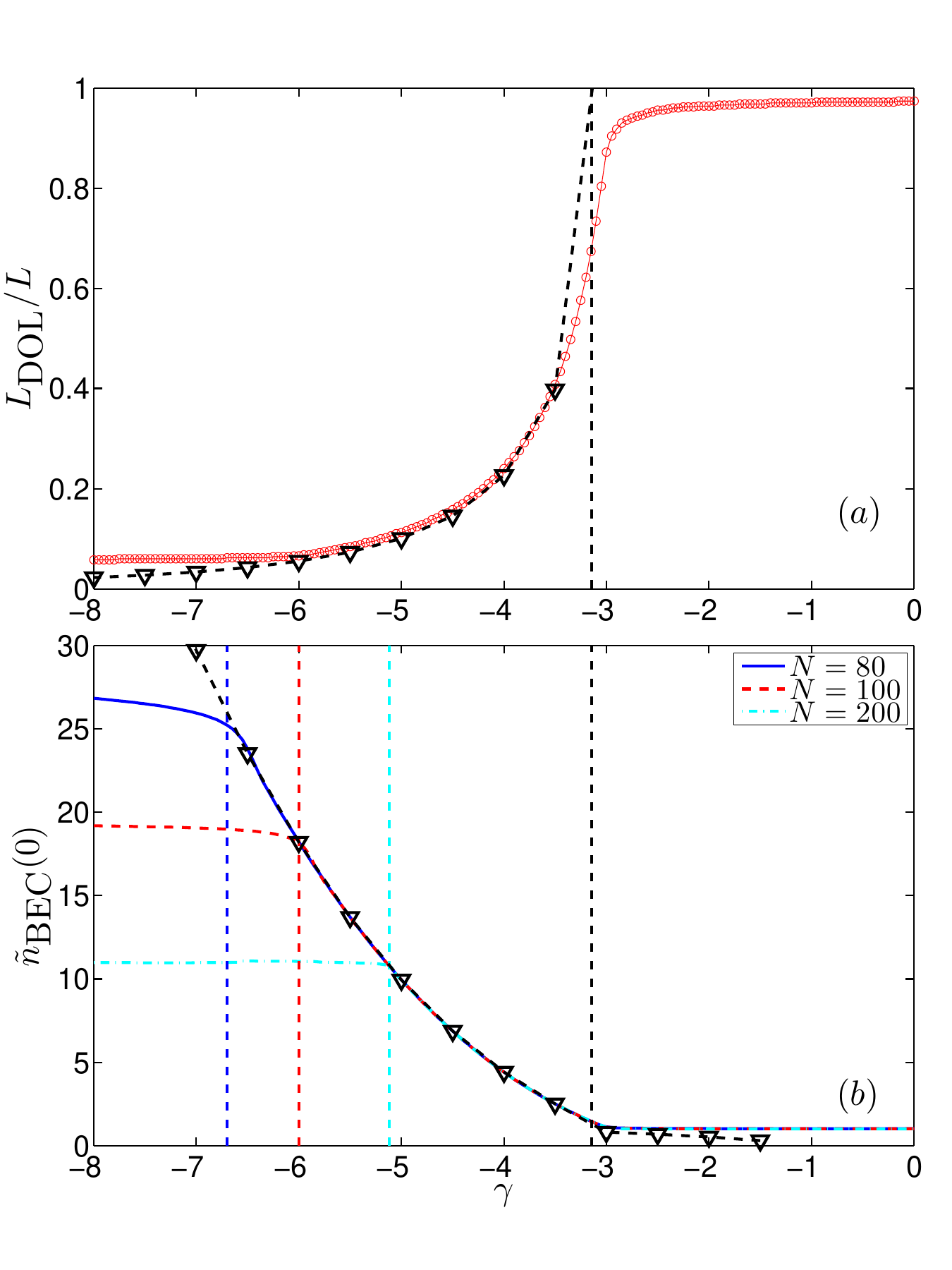} }
    \caption{(a) The rescaled density overlap length of the mixture as a function of the inter-species attraction strength, extracted from the abrupt change in the BEC density shown in Fig.~\ref{tgbec_att_nbec100ntg100L100_gbec1}(c).
    (b) The rescale central density of the BEC $ \tilde{n}_{BEC}(0)=n_{BEC}(0)/n_{F} $ as a function of the inter-species interaction strength, shown for three system sizes: $N =[80, 100, 200]$, with $\tilde{g}_{\text{BEC}} = 1$. The vertical black dashed line marks the critical point predicted by the mean-field (MF) theory. Black triangle dash lines are the results of quantum droplet mean-field method. } 
\label{tgbec_att_L100_gbec1_nbec0_line}
\end{figure}

\begin{figure}[htbp]
\centering
\subfigure{ \includegraphics[width=\linewidth]{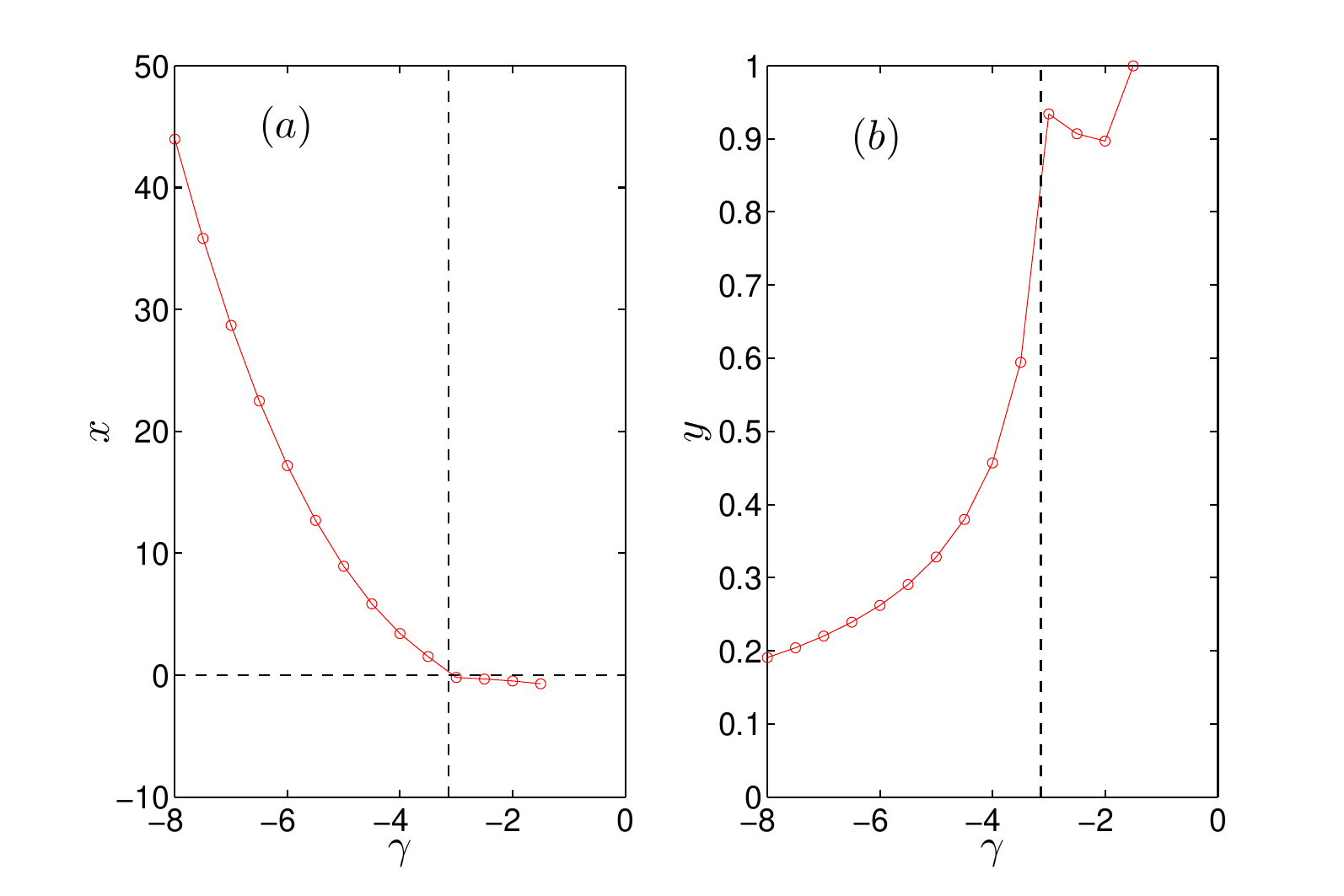} }
\caption{ Physical solutions to quantum droplet mean field equations. vertical dash lines are the critical points of mean field method. } 
\label{QD_MF_xy}
\end{figure}

\section{Quantum droplet}

The density distributions of the numerically obtained ground states of the BEC–TG mixture for four different values of the attractive interspecies interaction strength are shown in Fig.~\ref{tgbec_att_nbec100ntg100L100_gbec1}, with panel (a) displaying the BEC and panel (b) the TG gas. One can see that for no or weak attraction between the species ($\gamma= 0$ and $\gamma = -3$; blue and magenta lines, respectively), both components exhibit delocalized density profiles spanning the entire length of the box. However, for stronger attractions ($\gamma = -5$ and $\gamma = -8$; cyan and red lines), the BEC becomes increasingly localized near the center of the trap. Simultaneously, the TG gas density also accumulates in the central region, forming a co-localized structure with the BEC. Notably, both localised components exhibit a flat-top density profile, as previously discussed in the context of quantum droplets \cite{Astrakharchik}. This feature reduces the kinetic energy contribution while enhancing the negative binding energy of the system, thereby supporting the formation of a quasi self-bound state. The emergence of such a localized, flat-top structure is a characteristic signature of a quantum droplet in the BEC–TG mixture.

{\it Thomas-Fermi approximation--} 
To understand the properties of quantum droplet in BEC-TG mixture intuitively, let us assume that both species are uniformly distributed so that the kinetic energy of the BEC component can be neglected. In this approximation, only the  intra- and inter-component interaction energies need to be considered. The ground-state energy of a miscible BEC–TG mixture in this regime is then given by
\begin{equation}
    E_{0}=\frac{g_\text{BEC}N^{2}_\text{BEC}}{2L}+\frac{\pi ^{2}\hbar^2N^{3}_\text{TG}}{6mL^2}+g_\text{int}\frac{N_\text{BEC}N_\text{TG}}{L},
\label{E_MF_gcl}
\end{equation}
where the first term represents the standard mean-field interaction energy of the BEC \cite{pitaevskii_book}, the second term corresponds to the energy of a Tonks–Girardeau gas confined in a box of length $L$ \cite{Girardeau}, and the third term accounts for the interspecies interaction energy arising from the overlap of the two uniform density profiles. Introducing the energy density $\epsilon_{L}= E_0 / L$, which is a function of the densities $n_\text{BEC} = N_\text{BEC}/L$ and $n_\text{TG} = N_\text{TG}/L$, the stability of the system can then be assessed using the Hessian matrix of second derivatives of $\epsilon_{L}$ with respect to these densities. According to the standard thermodynamic criterion, the system is locally stable  with respect to variations in the component densities when the determinant of the Hessian matrix is positive \cite{Hessian_book}
\begin{equation}
    D(n_\text{BEC},n_\text{TG})=\det\left[ \frac{\partial^2 \epsilon_{L}}{\partial n_i \partial n_j}\right]>0,
\end{equation}
with $\{i,j\}\in\{\text{BEC},\text{TG} \}$. For the BEC–TG mixture described above, the determinant of the Hessian matrix can be calculated as $D(n_\text{BEC},n_\text{TG})=\pi^2 \hbar^2 g_\text{BEC}n_\text{TG}/m-g_\text{int}^2$,
which leads to the stability condition \cite{rakshit2019}
\begin{equation}
    \pi \hbar \sqrt{g_\text{BEC} n_\text{TG}/m} > |g_\text{int}|.
    \label{mf_gc}
\end{equation}
This inequality defines the threshold for the interspecies attraction strength, beyond which the system becomes unstable. In this regime, the two components can bind together and form a self-bound quantum droplet \cite{Petrov}.  It is worth noting that the above formula describes the mean-field contribution to energy only and may therefore  deviate from the result of exact numerical results.

In Figs.~\ref{tgbec_att_nbec100ntg100L100_gbec1}(c) and (d), we present the density distributions of the BEC and TG gas components over a large range of the inter-species interaction strength,  $\gamma$. As the attractive inter-species interaction increases, the initially delocalized density profiles of both components (green region) gradually transition into spatially localized distributions centered around the middle of the box (dark red region). The boundary marking this crossover is clearly identifiable in the plot. Interestingly, the TG gas retains a finite background density across the entire system, even at strong inter-species attraction, suggesting the coexistence of a localized droplet state with a delocalized TG component. To quantify the spatial extent of the overlapping region between the two components, we determine the density overlap length (DOL), $L_{\text{DOL}}$, by defining the quantum droplet boundary as the region where the rescaled BEC density exceeds the average background density of $0.01$ in Fig.~\ref{tgbec_att_nbec100ntg100L100_gbec1}(c).
From Fig.~\ref{tgbec_att_L100_gbec1_nbec0_line}(a) one can see that this quantity first decreases as the attraction increases, but saturates at strong coupling. It therefore provides a natural measure for distinguishing between different regimes of the system and we identify the regime where  $L_{\text{DOL}}$  varies strongly with  $\gamma$, approximately in the range  $\gamma \in [-6, -3]$, as the low-density quantum droplet phase. For stronger attraction ($\gamma \lesssim -6$), where the DOL becomes constant, we refer to the resulting state as a high-density quantum droplet. As shown by the horizontal magenta dots in Figs.~\ref{tgbec_att_nbec100ntg100L100_gbec1}(c) and (d), the critical point obtained from the mean-field criterion in Eq.~\eqref{mf_gc} closely aligns with the critical point indicated by solving the coupled equations, Eqs.~\eqref{eq_bec} and \eqref{eq_tg}. 

Since the TG gas is the limit of a Lieb-Liniger gas with infinite repulsion, it is worth noting that if only large but finite interactions (noting $g_\text{LL}\gg1$) are considered, higher-order corrections to the energy of the Bose gas appear \cite{Jiangyz}. In this case the ground state energy of quantum mixture is given by
\begin{align}          E_0=&\frac{g_\text{BEC}N^{2}_\text{BEC}}{2L}+\frac{g_\text{int}N_\text{BEC}N_\text{TG}}{L}\nonumber\\
&+\frac{\pi ^{2} \hbar^2 N^{3}_\text{TG}\left(1-4c_{0}/g_\text{LL}\right)}{6m L^2}+ \mathcal{O}\left(\frac{1}{g_\text{LL}^2}\right).
\end{align}
Here $g_\text{LL}$ is the intra-species interaction of the Lieb-Liniger Bose gas when it is close to the TG limit and $c_0$ is a constant. The above formula leads to a modified critical point given by
\begin{equation*}
     g_{c}\simeq - \pi \hbar \sqrt{g_\text{BEC}n_{\text{TG}}\left(1-\frac{4c_{0}}{g_{LL}}\right)/m}.
\end{equation*}
This shows that in the limit $1/g_\text{LL}\ll1$, the critical point shift only by a small amount and large but finite interactions should display TG regime effects.

As the system transitions from a delocalized gas mixture to a concentrated droplet state, the central BEC density increases significantly. This makes the BEC density at the center of the box,  $n_{\mathrm{BEC}}(x=0)$, another useful indicator of droplet formation. One can see from Fig.~\ref{tgbec_att_L100_gbec1_nbec0_line}(b) that, for fixed overall density, the central BEC density increases steadily within the low-density droplet regime and eventually saturates at a value dependent on the system size. Near the critical point, the curves for different system sizes collapse on top of each other, indicating that the onset of the quantum droplet follows a universal behavior consistent with a phase transition in the thermodynamic limit. Interestingly, the crossover from low-density to high-density quantum droplet shifts with system size, as indicated by the red, blue, and cyan dashed lines in Fig.~\ref{tgbec_att_L100_gbec1_nbec0_line}(b).  This indicates that the crossover is a smooth change rather than a sharp phase transition. A similar crossover behavior of Bose mixtures in 2D has recently been described in quantum Monte Carlo simulations of droplet formation by Spada {\it et al.}~\cite{spada_24} .

{\it Quantum droplet Thomas-Fermi analysis--} 
To more deeply understand these transitions we will next do an analysis similar to the one that led to Eq.~\eqref{E_MF_gcl} for the quantum droplet. After the phase transition to the quantum droplet state, we assume that the BEC and the TG have the respective particle numbers $N$ and $N_{1}$ in the droplet region of diameter $L_{q}$. The ground state energy of the droplet can then be calculated to be 
\begin{equation}
    E_{QD}=\frac{g_\text{BEC}N^{2}}{2L_{q}}+\frac{\pi ^{2}\hbar^2 N_{1}^{3}}{6 m L_{q}^2}+g_\text{int}\frac{N N_{1}}{L_{q}}+\frac{\pi ^{2} \hbar^2 (N-N_{1})^{3}}{6m(L-L_{q})^2},
    \label{eq:GSEnergyDroplet}
\end{equation}
where $L_q$ and $N_1$ are unknown and are determined by the extremization procedure using
$\partial E_{QD} /\partial L_{q} =0$ and  $\partial E_{QD}/ \partial N_{1} =0.$  
leads to the coupled equations
\begin{align}
\frac{g_{\text{BEC}}N^{2}}{2L_{q}^{2}}+\frac{\pi ^{2}\hbar^2 N_{1}^{3}}{3m L_{q}^3}+g_{\text{int}}\frac{N N_{1}}{L_{q}^{2}}=&\frac{\pi ^{2} \hbar^2 (N-N_{1})^{3}}{3m(L-L_{q})^3},
\label{Eq_qd1}\\
\frac{\pi ^{2} \hbar^2 N_{1}^{2}}{2mL_{q}^2}+g_{\text{int}}\frac{N}{L_{q}}=&\frac{\pi ^{2} \hbar^2 (N-N_{1})^{2}}{2m (L-L_{q})^2}.
\label{Eq_qd2}
\end{align}
For notational simplicity we introduce the parameterizations  $x=L/L_{q}-1$ and $y=N_{1}/N$, and find the two equations that determine the distribution of quantum mixture in quantum droplet region to be
\begin{eqnarray}
 && \frac{\pi^2 x^3 y^3}{3}+\frac{m g_\text{BEC}Lx^3}{2\hbar^2 N(1+x)}+\frac{m g_\text{int}Lyx^3}{\hbar^2 N(1+x)}=\frac{\pi^2 (1-y)^3}{3}, \label{qd_mf1}\\ 
 && \frac{\pi^2 x^2y^2}{2}+\frac{m g_\text{int}L}{\hbar^2 N}\frac{x^2}{1+x}=\frac{\pi^2 (1-y)^2}{2}. \label{qd_mf2}
\end{eqnarray}
It is worth noting that these equations only depend on the average density $N/L$, while the ground state energy, Eq.~\eqref{eq:GSEnergyDroplet}, has a more complex dependence on $L$ and $N$. 

Figure.~\ref{QD_MF_xy} shows the solutions to the algebraic Eqs.(\ref{qd_mf1}) and (\ref{qd_mf2}), which display nonanalytic singularity near the critical point. The solutions to these equations can be used directly to calculate the rescaled density overlap and the central BEC density, shown in  Fig.~\ref{tgbec_att_L100_gbec1_nbec0_line}(a)) and (b) as black dashed lines with triangles.  One can see good agreement in the low-density quantum droplet parameter regime, while the behaviour in the high-density quantum droplet parameter regime is less well captured. As the attraction strength increases, the mixture becomes self-bound inside the narrow quantum droplet region, which leads to the kinetic energy of BEC becoming more important and leading to the increasing disagreement between the red lines and the triangle lines in Fig.~\ref{tgbec_att_L100_gbec1_nbec0_line}(a).

It is also worth stressing that the critical point for droplet formation in our model depends explicitly on the density of the Tonks-Girardeau gas, and that the spatial extent of the droplet, $L_{\mathrm{DOL}}$, varies with the inter-species attraction.    These two features highlight distinct characteristics of the BEC–TG mixture when compared to the usual LHY-type quantum droplets formed in weakly interacting Bose-Bose mixtures \cite{Petrov,Petrov_liquid}. This distinction originates from the asymmetry in the intra-species interactions between the BEC and TG components, which allows for the formation of bound states with unequal numbers of particles from each species. Near the critical inter-species attraction strength, a large number of TG particles are required to bind all the BEC atoms, leading to a shallow bound state and a low-density droplet. As the inter-species attraction increases further, fewer TG particles are needed to bind the BEC component, resulting in a more compact and deeply bound state—a high-density droplet.

The density profiles and crossover points reported in our study can therefore serve as useful experimental benchmarks for the observation of low-to-high density droplet crossover in ultracold atomic gases.

 In the limit $N_{\text{TG}}\ll N$, where all TG atoms are part of the droplet, one has $N_1=N_{\text{TG}}$ and the ground state energy is then given by 
\begin{equation}
    E=\frac{g_{\text{BEC}}N^{2}}{2L_{q}}+\frac{\pi ^{2}\hbar^2 N_{\text{TG}}^{3}}{6m L_{q}^2}+g_{\text{int}}\frac{N N_{\text{TG}}}{L_{q}}.
\end{equation}
This allows one to calculate the extremal value  $\frac{\partial E}{\partial L_{q} }=0$ explicitly as
\begin{equation}
    L_{q}/L=\frac{\pi^2 \hbar^2 n_{\text{TG}}^3}{3m (|g_{\text{int}}|nn_{\text{TG}}-g_{\text{BEC}}n^{2}/2)},
\end{equation} 
where the average densities are defined as $n_{\text{TG}}=N_{\text{TG}}/L$ and $n=N/L$. One can see that within the deep attraction regime $|g_{\text{int}}|>g_{\text{BEC}}n / 2 n_{\text{TG}}$, it follows that the diameter of the quantum droplet varies with the attractive interaction strength $|g_{\text{int}}|$. Outside this parameter regime, the mixture tends to expand into the whole space as miscible phase.

\begin{figure}[tb]
\centering 
\subfigure{ \includegraphics[scale=0.28]{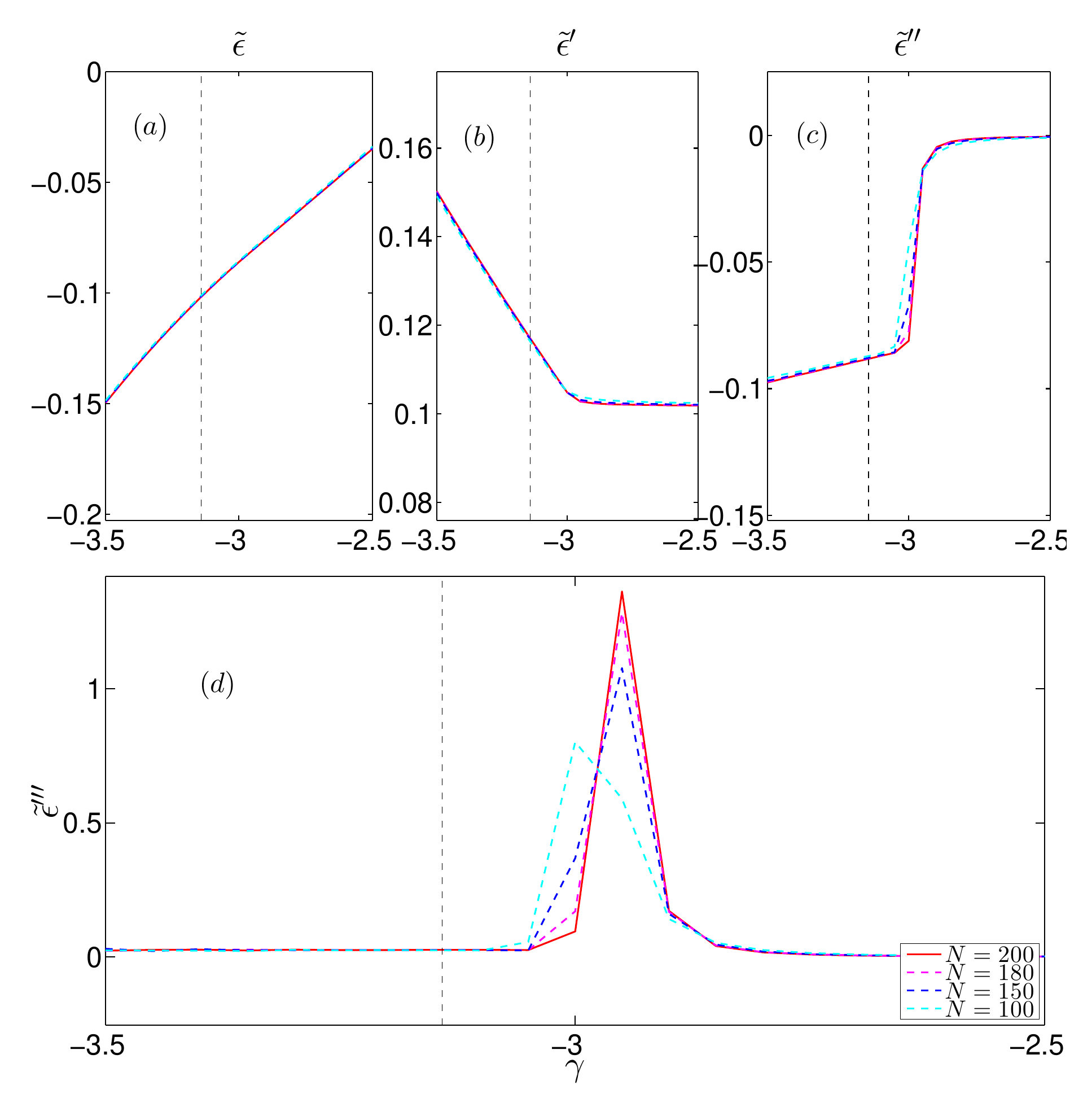} }
\caption{Ground state rescaled energy $\tilde{\epsilon}=E_{0}/N/E_{F} $ (a) and its first-order(b), second-order (c) and third-order (d) derivatives as a function of the rescaled interspecies attraction $\gamma$ near critical point. Here system size $[200,180,150,100]$. The black dashed line indicates the critical point obtained above. } 
\label{tgbec_att_LN_E0diff}
\end{figure}

\section{Third-order critical point} 
In the following we propose that the transition from a miscible mixture to a quantum droplet in our model constitutes a third-order phase transition, as defined in the framework of singularities in higher-order energy derivatives \cite{Gross_Edward,Eisele}. This claim is supported by the numerical results shown in Fig.~\ref{tgbec_att_LN_E0diff}, where we plot the ground-state  rescaled energy (a), along with its first(b),second (c) and third (d) order derivatives with respect to the inter-species interaction strength  $\gamma$. In Figs.~\ref{tgbec_att_L100_gbec1_nbec0_line} and~\ref{tgbec_att_LN_E0diff}, two distinct singular points can be observed. The first, marked near the black dashed line (which corresponds to the result obtained using the Thomas-Fermi approximation), is consistent across different system sizes and thus represents a true phase transition point. In contrast, the second singular point, which corresponds to the points indicated by blue, red, and cyan dashed lines in Fig.~\ref{tgbec_att_L100_gbec1_nbec0_line}(b), shifts with system size and corresponds to the smooth crossover to the high-density droplet.  

Crucially, the second-order derivative shown in subplot (c) remains continuous at the true critical point, while the third-order derivative shown in subplot (d) becomes singular. This behavior satisfies the criteria for a third-order phase transition \cite{Eisele}, which is noteworthy due to their relatively rare appearance in physical systems. Interesting examples include the Gross-Witten transition in quantum chromodynamics (QCD) \cite{Gross_Edward}, phase transitions in random matrix models \cite{majumdar2014top}, and transitions in Coulomb gases \cite{cunden2017}. Observing a third-order transition in the context of the BEC–TG mixtures represents a different mechanism for quantum droplet formation and stands in stark contrast to the more familiar first-order transition.

\begin{figure}[htbp]
\centering
\subfigure{ \includegraphics[scale=0.325]{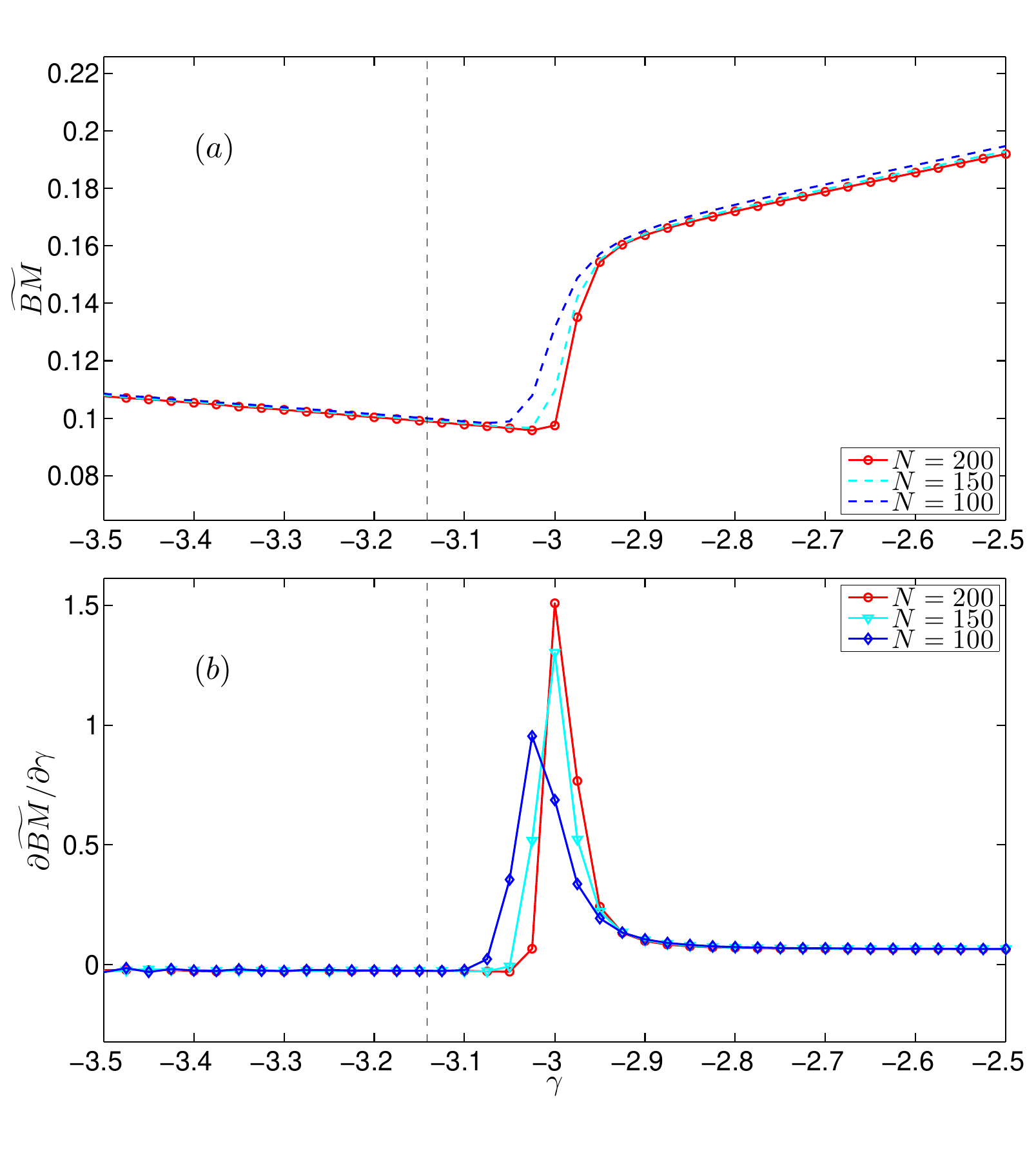} }
\caption{The rescaled bulk modulus $\widetilde{BM}=BM/E_{F}/k_{F}$  (a)  and its derivative (b) over attraction $\gamma$ near critical point vary with attraction for three system size $N=[200,150,100]$.  Black dash line is the mean-field critical point. Near critical point, bulk modulus is always the continuous function of attraction $\gamma_{\text{int}}$ and shows data collapse,  the derivative of bulk modulus become singular and the peaks height increase with system size. } 
\label{tgbec_att_BMdiff}
\end{figure}

The transition can also be understood by analyzing the compressibility of the gas phase of the miscible quantum mixture  and the liquid phase of quantum droplet. For the case where the  quasi-BEC and the TG gas have equal density $n$, the compressibility can be calculated as
\begin{equation}
    \kappa=\frac{1}{n} \frac{\partial n}{ \partial P}=\frac{1}{n^2(g_\text{BEC}+\hbar^2 \pi^2n /m+2g_\text{int})}.
\end{equation}
One can see that for a gas phase of a miscible quantum mixture with small $g_\text{int}$, the compressibility is finite and positive, while for strong attractions, $g_\text{int}<-(g_\text{BEC}+\hbar^2 \pi^2n/m)/2$, the compressibility changes its sign. However, in the quantum droplet phase, the formula for the compressibility above becomes invalid, because the density of the quantum mixture becomes inhomogeneous in space as visible in Fig.~\ref{tgbec_att_nbec100ntg100L100_gbec1}. In order to circumvent this limitation, we compute the bulk modulus (BM), which applies to the gas phase and the liquid phase given by
\begin{equation}
    BM=-L \frac{\partial P}{\partial L}.
\end{equation}
This quantity characterizes the incompressibility or elasticity of matter (see \cite{Bulkmodulus}) and the pressure is given by $P=-\partial E/ \partial L$.  From Fig.~\ref{tgbec_att_BMdiff}(a) one can see that the BM is a continuous function of the rescaled attractive interaction $\gamma$, without a jump or a diverging singularity. One can also see from Fig.~\ref{tgbec_att_BMdiff}(a) that the BM  data collapses onto one line for three system sizes $N=[200,150,100]$ with finite-size effects at critical point. Since the bulk modulus is a second-order quantity and a continuous function of the attraction strength $\gamma$, this therefore excludes a second-order phase transition. One can also see that the BM has local minimum near the critical point, which separates two different tendencies of the elasticity of the matter: one applies in the gas phase of the miscible quantum mixture and the other in the liquid phase of the quantum droplet. Finally,  Fig.~\ref{tgbec_att_BMdiff}(b) shows the derivative of the bulk modulus as a function of $\gamma$ and one can see that this third-order quantity becomes singular as the peaks increase with system size. The derivative of the bulk modulus also changes its sign when crossing the critical point. These results strongly support a third-order transition from a gas quantum mixture to a quantum droplet.

\section{Stability }
The stability of quantum droplets is an important point in both theoretical and experimental studies, particularly regarding the question of whether a droplet-like mixture will collapse when more particles are added. A stable quantum droplet requires the ground-state energy to exhibit (at least local) convexity as a function of particle density, such that the chemical potential transits from negative to positive values as the density increases. Physically, a negative chemical potential ensures binding of the system, while a positive chemical potential implies that adding more particles increases the energy, thereby preventing collapse and stabilizing the droplet. In the case of conventional LHY-stabilized droplets \cite{Petrov}, the energy density includes a repulsive  $n^{5/2}$  term from quantum fluctuations and an attractive  $n^2$  mean-field contribution, which together provide a stability mechanism.

\begin{figure*}[tb]
\centering
\subfigure{ \includegraphics[scale=0.39]{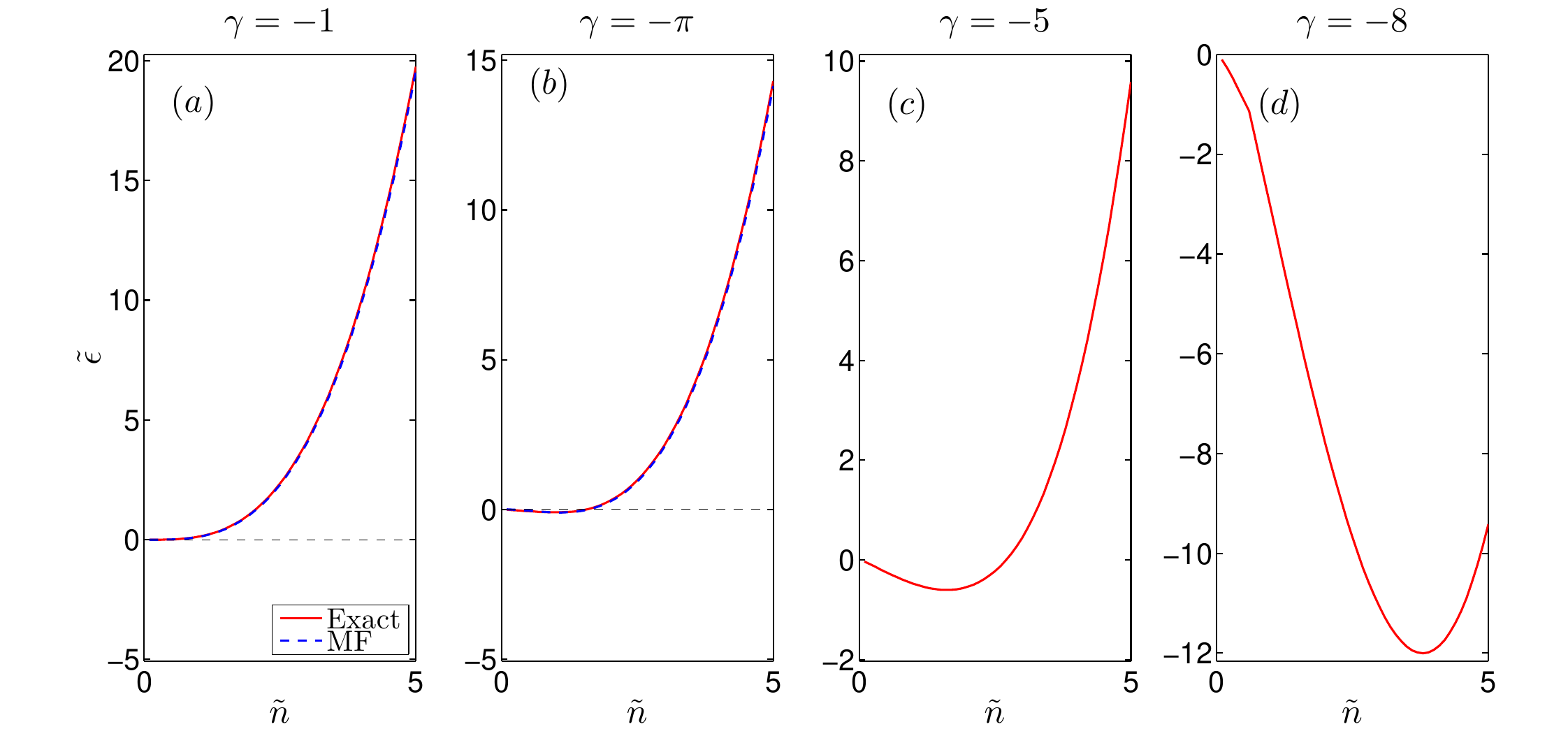} }
\caption{Ground state rescaled energy $\tilde{\epsilon}$ as a function of the rescaled gas density  $\tilde{n}$ for four rescaled attraction $\gamma$.  }  
\label{att_E0rho_gint}
\end{figure*}

To assess stability in the model presented here, we consider the balanced-density case  $n_{\text{TG}} = n_{\text{BEC}} = n$. The mean-field energy density in the Thomas-Fermi limit for this scenario is given by
\begin{equation}
    \frac{E_{0}}{L}=\frac{g_\text{BEC}n^{2}}{2}+\frac{\pi^{2} \hbar^2 n^3}{6m}+g_\text{int}n^{2},
    \label{E0_dens}
\end{equation}
which includes the BEC self-interaction, the kinetic energy of the Tonks-Girardeau gas, and the inter-species coupling term. Figure.~\ref{att_E0rho_gint} shows a comparison between the numerically exact energy density (red lines) and the Thomas-Fermi limit predictions (blue dashed lines) for different values of  $\gamma$. For  $\gamma$  above or close to the critical point (see panels (a) and (b)), the energy density increases monotonically with particle density, indicating a miscible, unbound state. In this regime, the Thomas-Fermi results agree well with the full numerical results. However, at  $\gamma = -5$ (panel (c)), the energy density initially decreases with increasing  $n$ , signaling the formation of a bound state between the BEC and TG gases. As  $n$  increases further, the repulsive cubic term  $\propto n^3$  balances the attractive quadratic term  $\propto n^2$, leading to a local energy minimum at finite density. This non-monotonic behavior indicates the emergence of a stable quantum droplet. For even stronger attraction, such as  $\gamma = -8$  (panel (d)), the local minimum deepens, corresponding to the formation of a high-density quantum droplet. These results confirm that the quantum droplet formed in the BEC–TG mixture is stable and that its stabilization arises from the interplay between attractive inter-species interactions and the kinetic pressure of the TG gas, analogous, but not identical, to the LHY stabilization mechanism in conventional Bose-Bose mixtures.

\section{ Conclusion }
By solving for the ground state of a density-coupled one-dimensional quasi Bose-Einstein condensate (BEC)–Tonks-Girardeau (TG) gas mixture, we have demonstrated the formation of a quantum droplet in this system. We have identified low-density and high-density quantum droplet regimes and characterized the crossover between them, using indicators such as the density overlap length and the central density of the BEC component within the droplet. Our analysis shows that the transition from a miscible mixture to a quantum droplet in the BEC–TG system constitutes a third-order phase transition, in stark contrast to the first-order transition observed in conventional LHY-stabilized quantum droplets in Bose–Bose mixtures. 

In future work, it would be interesting to investigate the collective excitations~\cite{huliu_pra,Xucong} and sound velocity of the BEC–TG mixture~\cite{ chin_vs, Natale}, and to explore the nonequilibrium quench dynamics of the system to further elucidate the properties and formation mechanisms of quantum droplets. Our work shows that it is feasible to simulate and study  higher order phase transition with ultracold atomic systems, which may extend the understanding about phase transition frameworks.
 
\section{acknowledgments}
This work was supported by the Okinawa Institute of Science and Technology Graduate University and utilized the computing resources of the Scientific Computing and Data Analysis section of the Research Support Division at OIST. We also acknowledge support from the JST Grant No.~JPMJPF2221 and the JSPS Bilateral Program No.~JPJSBP120244202. This work was also supported by National Key Research and Development Program of China (Grant No. 2021YFA0718304), by  the NSFC (Grants No.12574295).

\bibliography{qdroplet}

\end{document}


\title{Quantum droplets in one-dimensional mixtures of quasi Bose-Einstein condensates and Tonks-Girardeau gases}  

\author{Wen-Bin He}
\email{wenbin.he@oist.jp}
\affiliation{Quantum Systems Unit \& Okinawa Center for Quantum Technologies, Okinawa Institute of Science and Technology Graduate University, 904-0495 Okinawa, Japan}%

\author{Su Yi}
\email{syi@itp.ac.cn}
\affiliation{Institute of Fundamental Physics and Quantum Technology \& School of
Physical Science and Technology, Ningbo University, Ningbo, 315211, China}
\affiliation{Peng Huanwu Collaborative Center for Research and Education, Beihang University, Beijing 100191, China}


\author{Thomas Busch}%
\email{thomas.busch@oist.jp}
\affiliation{Quantum Systems Unit, Okinawa Institute of Science and Technology Graduate University, 904-0495 Okinawa, Japan}%

\date{\today}

\begin{widetext}

\begin{center}
    \textbf{Supplemental Material: Quantum droplets in one-dimensional mixtures of quasi Bose-Einstein condensates and Tonks-Girardeau gases }
\end{center} 
\begin{center}
Wen-Bin He, Su Yi, and Thomas Busch
\end{center} 


In this supplemental material, we will present more details supporting the results presented in the  main text, including the derivation of the equations of motion, extra results for balanced and imbalanced density cases, and the Thomas-Fermi analysis in mixture and quantum droplet phase.

\section{Equations of motion}

Taking advantage of the Bose-Fermi mapping theorem, the energy functional of a BEC-TG gas mixture in a box is given by
\begin{equation}
    E=\int dx  \psi^{*}(x) \left[-\frac{1}{2}\frac{\partial^2}{\partial x^2}+\frac{1}{2}g_{\text{BEC}}|\psi|^2\right]\psi(x)+\sum_{n=1}^{N_{\text{TG}}}\phi_n^{*}(x)\left(- \frac{1}{2}\frac{\partial^2}{\partial x^2}\right)\phi_n(x)+  g_{\text{int}}|\psi(x)|^2 \rho(x),
    \label{eq_E}
\end{equation}
where the normalization constraint on the mean-field wave function of the BEC $\psi(x)$ and the single particle eigenstates for the TG atoms $\{ \phi_n(x) \}$ are given by
\begin{eqnarray}
 \int dx |\psi(x)|^2=N_{\text{BEC}}\quad\text{and}\quad  \int dx |\phi_n(x)|^2 =1.
\end{eqnarray}
Using the Lagrange multiplier method we get the action functional
\begin{equation}
    S=\int \mathcal{L}(x) dx=E-\mu \int dx |\psi(x)|^2- \sum_{n}\epsilon_{n} \int dx |\phi_n(x)|^2,
\end{equation}
which should be extremised by the wave functions $\psi(x)$ and $\{ \phi_n(x) \}$ as
\begin{equation}
    \frac{\delta \mathcal{L}(x)}{\delta \psi^{*}(x)}=0\quad\text{and}\quad \frac{\delta \mathcal{L}(x)}{\delta \phi_n^{*}(x)}=0.
\end{equation}
Doing this, one arrives at the equations of motion for the mixture
\begin{align} 
    \label{eq:BEC}
    \left(-\frac{1}{2}\frac{\partial^2}{\partial x^2}+ g_{\text{BEC}}|\psi|^2 +g_{\text{int}}\rho(x)\right)\psi(x)=\mu\psi(x), \\
    \label{eq:TG}
    \left(-\frac{1}{2}\frac{\partial^2}{\partial x^2}+g_{\text{int}} |\psi(x)|^2\right)\mathbf{\Phi(x)}=\epsilon \mathbf{\Phi(x)},
\end{align}
where the many-body TG wavefunction is given by $\mathbf{\Phi}(x_1,\cdots,x_N)=\prod_{i<j}\sgn(x_i-x_j) \det_{m,n}[\phi_{m}(x_{n})]/\sqrt{N_\text{TG}!}$. The density distribution of TG is given as $\rho(x) = \sum_{n=1}^{N_{\text{TG}}} |\phi_n(x)|^2$.

\section{Thomas-Fermi analysis}

The ground state energy of the quasi-BEC-TG gas systems in the miscible phase and in the Thomas-Fermi limit is given by
\begin{equation}
    E_{0}=\frac{g_{\text{BEC}}N^{2}_{\text{BEC}}}{2L}+\frac{\pi ^{2}N^{3}_{\text{TG}}}{6L^2}+g_{\text{int}}\frac{N_{\text{BEC}}N_{\text{TG}}}{L}.
\end{equation}
Introducing the energy density as $\epsilon=E_{0}/L$, one can see that it is only a function of the atomic densities, $n_{\text{BEC}}$ and $n_{\text{TG}}$, and we show this quantity in Fig.~\ref{mf_E0_contour}. One can see that with increasing interspecies attraction $g=-[1,\pi,5,8]$, the energy surface form local mimum from  monotonic surface, which imply the phase instability and bound state.


\begin{figure}[tb]
    \includegraphics[width=\linewidth]{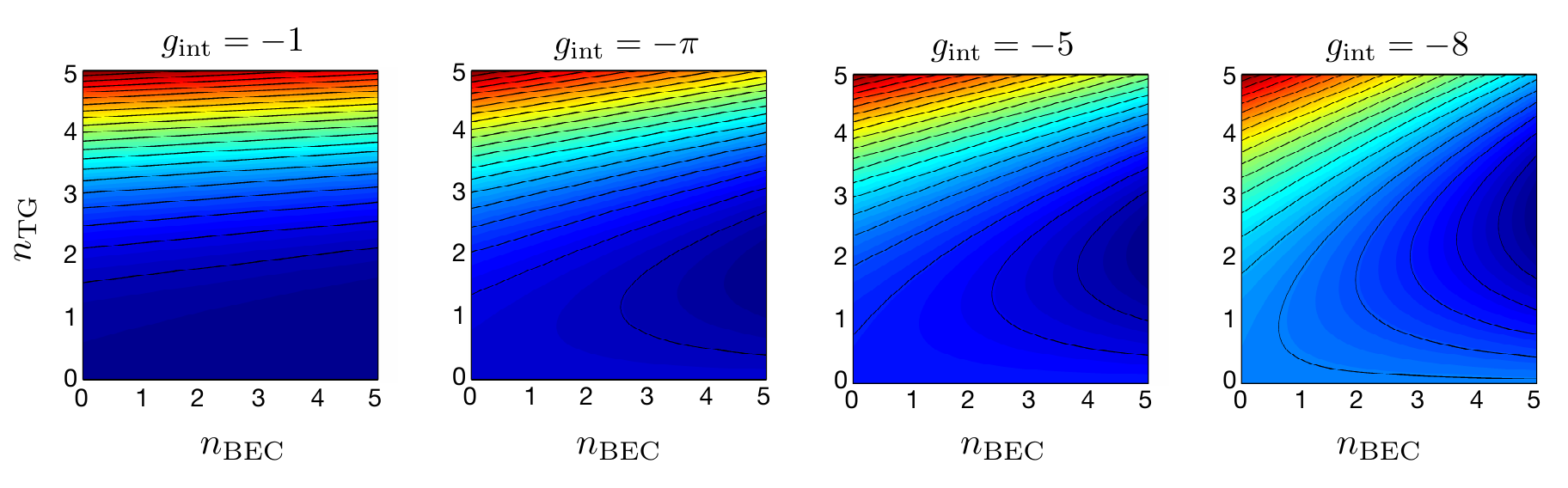} 
    \caption{The energy density for the two component system in the Thomas-Fermi limit as function of the densities of the BEC and the TG gas for different interspecies attractions $g_{\text{int}} = -[1,\pi,5,8]$.  The black lines are the contour lines of fixed energy density.} 
\label{mf_E0_contour}
\end{figure}

The stability of the miscible system can be determined by confirming the positivity of the determinant of the Hessian matrix of the energy density 
\begin{equation}
    D(n_{\text{BEC}},n_{\text{TG}})=\det\left[ \frac{\partial^2 \epsilon}{\partial n_{i} \partial n_{j}}\right],
\end{equation}
where $\{i,j\}\in\{\text{BEC},\text{TG}\}$. This gives
\begin{equation}
    D(n_{\text{BEC}},n_{\text{TG}})=
    \begin{vmatrix}
     g_{\text{BEC}} & g_{\text{int}}\\
     g_{\text{int}} & \pi^2 n_{\text{TG}} 
    \end{vmatrix}
    =\pi^2 g_{\text{BEC}}n_{\text{TG}}-g_{\text{int}}^2.
\end{equation}
The system is therefore stable if 
\begin{equation}
\pi \sqrt{g_{\text{BEC}}n_{\text{TG}}}>|g_{\text{int}}|,
\end{equation} 
and otherwise a transition to another phase has to occur. For our systems this is where the quantum droplet emerges.




Once the system is in a regime where the quantum droplet is formed, a similar analysis as above can be performed by assuming that a fixed number of BEC and TG gas particles, $N$ and $N_{1}$, reside inside the quantum droplet region of size $L_{q}$. The droplet ground state energy can then be written as
\begin{equation}
    E_{ \text{QD}}=\frac{g_{\text{BEC}}N^{2}}{2L_{q}}+\frac{\pi ^{2}N_{1}^{3}}{6L_{q}^2}+g_{\text{int}}\frac{N N_{1}}{L_{q}}+\frac{\pi ^{2}(N-N_{1})^{3}}{6(L-L_{q})^2},
\end{equation}
and one can see that it depends on two variables, $(L_{q}$ and $N_{1})$. Calculating the extremal values 
\begin{equation}
    \frac{\partial E_{\text{QD}}}{\partial L_{q} }=0,
    \quad \text{and}\quad
    \frac{\partial E_{\text{QD}}}{\partial N_{1} }=0
\end{equation}
leads to the coupled equations
\begin{align}
\frac{g_{\text{BEC}}N^{2}}{2L_{q}^{2}}+\frac{\pi ^{2}N_{1}^{3}}{3L_{q}^3}+g_{\text{int}}\frac{N N_{1}}{L_{q}^{2}}=&\frac{\pi ^{2}(N-N_{1})^{3}}{3(L-L_{q})^3},
\label{Eq_qd1}\\
\frac{\pi ^{2}N_{1}^{2}}{2L_{q}^2}+g_{\text{int}}\frac{N}{L_{q}}=&\frac{\pi ^{2}(N-N_{1})^{2}}{2(L-L_{q})^2}.
\label{Eq_qd2}
\end{align}
To solve these equations, we parameterize the existing variable as $x=L/L_{q}-1$ and $y=N_{1}/N$, which leads to 
\begin{align}
\frac{\pi^2 x^3 y^3}{3}+\frac{g_{\text{BEC}}L}{2N}\frac{x^3}{1+x}+\frac{g_{\text{int}}L}{N}\frac{yx^3}{1+x}=&\frac{\pi^2 (1-y)^3}{3},\\
\frac{\pi^2 x^2y^2}{2}+\frac{g_{\text{int}}L}{N}\frac{x^2}{1+x}=&\frac{\pi^2 (1-y)^2}{2}.
\end{align}
The immediately noticeable trivial solution of $x=0$ and $y=1$ corresponds to the miscible phase. To find other solutions, we numerically solve the coupled equations under the constraints $x>0$ and $0<y<1$. The solutions are shown in Fig.~\ref{QD_MF_xy}. From both subplots one can see that there is non-differential singularity of phase transition near critical point. The diameter of quantum droplet resulting from this method is compared with exact numerical results in Fig.~2(a) in the main text. The results of this approach agree qualitatively well with the exact numerical results.
Let us finally consider the \emph{extreme case} where all TG atoms are part of the droplet, $N_1=N_{\text{TG}}$, while $N_{\text{TG}}\ll N$ . The ground state energy is then given by 
\begin{equation}
    E=\frac{g_{\text{BEC}}N^{2}}{2L_{q}}+\frac{\pi ^{2}N_{\text{TG}}^{3}}{6L_{q}^2}+g_{\text{int}}\frac{N N_{\text{TG}}}{L_{q}},
\end{equation}
and calculating the extremal value $\frac{\partial E}{\partial L_{q} }=0$ leads to
\begin{equation}
    \frac{g_{\text{BEC}}N^{2}}{2L_{q}^{2}}+\frac{\pi ^{2}N_{\text{TG}}^{3}}{3L_{q}^3}=|g_{\text{int}}|\frac{N N_{\text{TG}}}{L_{q}^{2}}.
\end{equation}
This results in
\begin{equation}
    L_{q}/L=\frac{\pi^2N_{\text{TG}}^3}{3(|g_{\text{int}}|NN_{\text{TG}}-g_{\text{BEC}}N^{2}/2)L}=\frac{\pi^2n_{\text{TG}}^3}{3(|g_{\text{int}}|nn_{\text{TG}}-g_{\text{BEC}}n^{2}/2)},
\end{equation} 
where the average densities are defined as $n_{\text{TG}}=N_{\text{TG}}/L$ and $n=N/L$. This means that the system is stable if $|g_{\text{int}}|>g_{\text{BEC}}n / 2 n_{\text{TG}}$, showing that the diameter of the quantum droplet  varies with attractive interactions $|g_{\text{int}}|$. 

 \vspace{2em}

\begin{figure}[htbp]
\centering
\includegraphics[width=\linewidth]{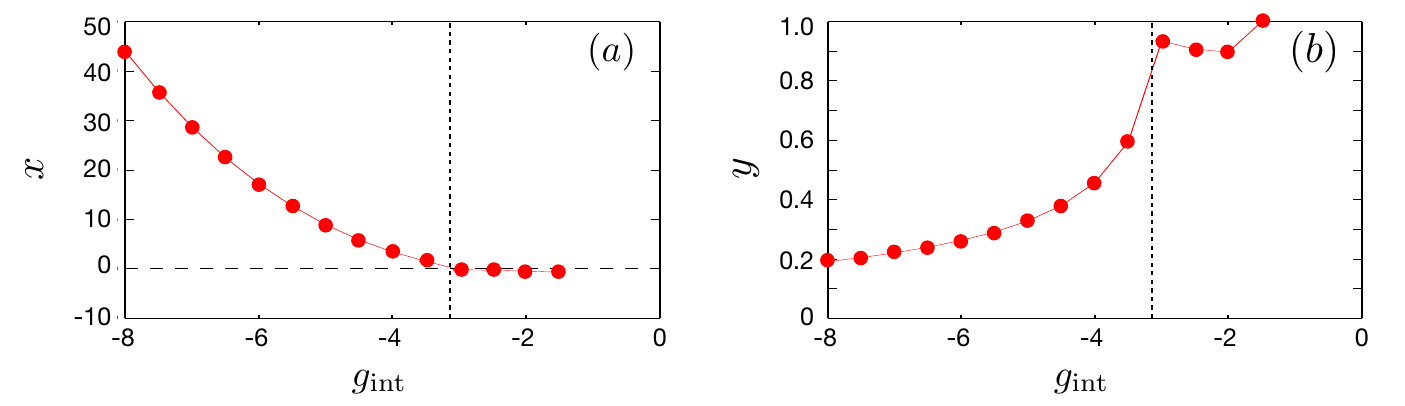}
\caption{ Physical solutions to quantum droplet mean field equations. vertical dash lines are the critical points of mean field method. } 
\label{QD_MF_xy}
\end{figure}

\end{widetext}